\documentclass{acs_arxive_nature}

\usepackage{caption}
\usepackage{graphicx}
\usepackage{rotating}

\newcommand{\gtsim}{\raisebox{-1.0ex}{$\stackrel{\textstyle>}{\sim}$}}
\newcommand{\ltsim}{\raisebox{-1.0ex}{$\stackrel{\textstyle<}{\sim}$}}
\def\kms{km~s$^{-1}$}

\def\hinode{{\sl Hinode}}

\def\p78{{\sl P78-1}}

\def\sdo{{\sl SDO}}

\def\kms{km~s$^{-1}$}

\def\etal{{\it et~al.}}

\newcounter{iref}
\title{Small-scale filament eruptions as the driver of solar coronal hole X-ray jets}

\author{Alphonse C.~Sterling$^1$, Ronald L. Moore$^{2,1}$, David A. Falconer$^{2,1}$, \& Mitzi Adams$^1$}

\begin{document}

\maketitle

\begin{affiliations}
 \item NASA/Marshall Space Flight Center, Huntsville, Alabama 35812
 \item Center for Space Plasma and Aeronomic Research, University of Alabama in Huntsville, Huntsville, AL 35899
\end{affiliations}

\begin{abstract}

{\boldmath

Solar X-ray jets are evidently made by a burst of reconnection of closed magnetic field in a jet's base with ambient
``open'' field\cite{shibata.et92,cirtain.et07}.  In the widely-accepted version of the ``emerging-flux'' model, that
reconnection occurs at a current sheet between the open field and emerging closed field and also makes a compact hot
brightening that is usually observed at the edge of the jet's base\cite{shibata.et92,yokoyama.et95}.  Here we report on
high-resolution X-ray and EUV observations of 20 randomly-selected X-ray jets in polar coronal holes.  In each jet,
contrary to the emerging-flux model, a miniature version of the filament eruptions that initiate coronal mass ejections
(CMEs)\cite{hirayama74,shibata.et95,moore.et01,chen11} drives the jet-producing reconnection, and the compact hot
brightening is made by internal reconnection of the legs of the minifilament-carrying erupting closed field, analogous to
solar flares of larger-scale eruptions.   Previous observations have found that some jets are driven by base-field
eruptions\cite{moore.et77,sterling.et10,raouafi.et10,nistico.et09}, but only one such study, of only one jet,
provisionally questioned the emerging-flux model\cite{adams.et14}.  Our observations support the view that solar filament
eruptions are made by a fundamental explosive magnetic process that occurs on a vast range of scales, from the biggest
CME/flare eruptions down to X-ray jets, and perhaps down to even smaller jets that are candidates for powering coronal
heating\cite{depontieu.et11,moore.et13,raouafi.et10}.  A picture similar to that suggested by our observations was drawn
before, inferred from different observations and based on a different origin of the erupting minifilament flux rope\cite{shibata99} (see Methods).

}  

\end{abstract}

Solar X-Ray jets are imaged from space in the $\sim$0.2---2.0~keV range.  They are dynamic (upward
velocities $\sim$200~\kms), long ($\sim$5$\times 10^4$~km), narrow ($8\times 10^3$~km), and transient (lifetimes
$\sim$10~min)\cite{shimojo.et96,savcheva.et07}.   In the commonly-accepted version of the emerging-flux
model\cite{yokoyama.et95,nishizuka.et08,moreno-insertis.et13,archontis.et13,fang.et14}, an emerging bipole enters
a dominant-polarity (say, negative) ambient open  field, and the bipole's minority-polarity (positive) side can
reconnect with coronal field  at the location of the magnetic null region between the bipole and the ambient
field.  In this model, a burst of reconnection connects the outside of the bipole with adjacent coronal field, producing a
small loop on the outside of the emerging bipole's minority-polarity foot, and reconnects open field
to the outside of the bipole's majority-polarity foot.  An X-ray jet  develops as 
reconnection-heated material flows out along the new open field strands.  Additionally, the reconnection-formed
small loop at the emerging field's edge is the model's explanation for the observed base-edge compact X-ray jet bright 
point (JBP\@). In a later-suggested extension of the emerging-flux model, the emerged bipole explodes as it
reconnects,  forming a ``blowout jet'' with a relatively-broad spire\cite{moore.et13}.  (See Methods and
Extended Data Fig.~1 for details of the emerging-flux model.)

To assess observationally the production of X-ray jets, we analysed 20 jets (Extended Data Table~1) in the solar
polar regions using X-ray images from the X-ray telescope (XRT) on the \hinode\ satellite\cite{kosugi.et07}, which
detects a broad temperature range of coronal plasmas hotter than about 1.5~MK\@.  We used concurrent EUV images
from the Solar Dynamics Observatory's (\sdo) Atmospheric and Imaging Assembly (AIA)\cite{lemen.et12}, whose
various filters detect plasmas primarily over narrow temperature ranges centred at, e.g.\ $T \approx 0.05, 0.6, 1.6, \rm{or}\
2.0 $~MK respectively for wavelengths of 304~\AA, 171~\AA, 193~\AA, 211~\AA\@.  (See Methods.)

Figure~1 shows a typical example of our results in both soft X-ray (Figs.~1a---1c) and EUV (Figs.~1d---1f)
images.  Between Figures~1a  and~1b the jet's spire, arched base, and JBP all begin brightening. Later
(Fig.~1c) the spire extends higher, with the JBP positioned about $10''$ west of the spire.  From a movie
constructed from the XRT images (see Extended Data Video~1abc), the JBP brightening  starts at $\sim$22:07~UT,
with the spire becoming visible $\sim 2.5$~min later.  Thus one can imagine this jet obeyed the emerging-flux
model, where {\it external reconnection} (i.e., reconnection occurring on the outside of the  closed driving
field\cite{sterling.et01}) of emerging field forms the JBP and gives rise to the spire at a displaced location.
Observing the same feature in AIA 193~\AA\ EUV images (Fig.~1, and in Extended Data Video~1def) however does
{\it not} support this interpretation.  These images clearly show a dark feature, similar to a small-scale solar
chromospheric filament (hereafter, ``minifilament''), moving upward and laterally, starting from
$\sim$22:06~UT\@.  Its velocity is $\sim 40$~\kms\ between 22:07~UT and $\sim$22:10~UT, when it reaches the apex
of the illuminated arched base of the X-ray jet.   After 22:10~UT, the minifilament is expelled in the spire of an
EUV jet that is the counterpart to the XRT jet.  In EUV the jet has both emission and absorption components,
with the minifilament evolving into part of the jet.  Significantly however, the JBP, both in soft X-rays and in
EUV, is at the location from where the minifilament erupted. Thus the JBP is the analog to the commonly-observed
solar flare arcade forming in the wake of larger-scale filament eruptions; such flare arcades are made by {\it
internal reconnection} (i.e., reconnection occurring on the inside of the closed driving
field\cite{sterling.et01}) of the legs of the erupting closed field of a filament.  This is {\it not} consistent
with the JBP resulting from external reconnection as proposed in the emerging-flux model.


We found an erupting minifilament to be discernible in AIA images of all 20 of the jets, with the  minifilament's
eruption starting near the location of the JBP\@. In most cases we could see that the JBP occurred where the
minifilament (or part of the minifilament) had been rooted in the surface prior to ejection; we could not verify
this arrangement in a few cases where the minifilament and JBP were along the same  line-of-sight,  but even then
the observations are consistent with the JBP occurring at the location from where the minifilament was ejected.  
Typically, first the minifilament starts to lift off from the surface, and then the JBP starts to brighten.  This
is similar to the situation with large-scale filament eruptions, where the eruption start precedes the
flare-brightening onset\cite{sterling.et05}.   Other than size scale, the eruptions of minifilaments in the
production of X-ray jets are indistinguishable from the commonly-observed eruptions of larger filaments in the
onsets of solar flares.  In some  cases (in Extended Data Table~1, events 4, 9 and 13, and maybe event 1), rather
than the entire minifilament lifting off, there is a whipping-like motion, with the JBP (flare) occurring below
the whipping minifilament or at the location where the fastest-moving part of the minifilament first detaches from
the solar surface. Thus all cases are consistent with the JBP being a small flare arcade forming in the wake of
the  erupting minifilament\cite{hirayama74,shibata.et95,moore.et01,chen11}.

We measured the plane-of-sky sizes and velocities of the minifilaments, during the period after they started to
erupt but prior to reaching the jet-spire location.  The average length of the
minifilaments  were $11''$ ($=8\times 10^3$~km) with a standard deviation of $4''.$  This is much smaller than
the sizes quoted for filaments from an extensive survey\cite{bernasconi.et05} ($3\times 10^4$---$1.1\times
10^5$~km), justifying the term ``minifilaments''.  (Perhaps-identical minifilaments have been identified on the
solar disk\cite{wang.et00}.)  Our measured average minifilament size is equal to the average width of X-ray
jets\cite{savcheva.et07}, consistent with the jets being driven by the minifilament eruptions.  We obtained mean
velocities and a standard deviation for the erupting minifilaments of $31 \pm 15$~\kms.  In all cases the true
sizes and speeds should tend to be larger than these plane-of-sky values.

X-ray jets have been classified as ``standard'' or ``blowout'' based on the morphology of the spire and the  intensity of
the rest of the jet's base compared to the JBP intensity:  A standard jet has a narrow spire with a relatively dim base,
while a blowout jet has a broad spire and a base that becomes about as bright as the JBP\cite{moore.et13}.  The
emerging-flux model suggests that the difference occurs depending on whether the emerging-flux structure remains largely
inert (standard jet), or erupts (blowout jet) as the jet forms.  Our new view is different: In a previous
study\cite{moore.et13} of our 20 events, we morphologically  classified 14 as blowout, five as standard, and one as
ambiguous.  We now find however that all 20 appear to  form the same way: from erupting filaments. A jet has blowout-jet
morphology if  the erupting filament strongly ejects from the base region (corresponding to an ejective larger-scale solar
eruption\cite{moore.et01}).  Standard-jet morphology seems to result when the erupting minifilament mainly does not escape
the closed-field base (maybe corresponding to confined larger-scale filament eruptions\cite{moore.et01}), or perhaps if
the eruption is ejective but very weak.  We envision that  there is a continuum of morphological jet types, likely
depending on the eruption's strength and whether the  erupting filament escapes the base.

From our observations we infer the schematic picture of Figure~2 for jet production.  Initially (Fig.~2a)
two bipoles sit side-by-side, the larger one corresponding to what we usually observe as the base of the  jet
(cf.~Fig.~1).  The smaller bipole contains substantial free energy in sheared and twisted magnetic field; that
field holds a minifilament. As with the case of large-scale solar eruptions, this field  becomes unstable by some
process; it then erupts outward, guided between the large bipole and the ambient open field. After the
minifilament's liftoff, reconnection occurs among the distended legs of the minifilament field (Fig.~1b), making a
``flare-arcade'' JBP via internal reconnection occurring inside the erupting field.   The spire starts as soon as
the outer envelope of the minifilament-carrying erupting field starts external reconnection with open field on the far side
of the large bipole. External reconnection continues and soon reconnects the field threading the  erupting
minifilament with far-side open field, injecting minifilament plasma along that open field.  The external
reconnection also adds a new hot layer to the larger bipole (larger red loop in Fig.~2c).

If the erupting minifilament-carrying field blows out beyond the large bipole's apex (Figs~2b---2c), then
widespread external reconnection results; this creates a broad jet spire characteristic of blowout jets.  If the
erupting field stalls near the apex of the large bipole  (and/or if the eruption is weak enough), the external
reconnection produces only a narrow jet, characteristic of a standard jet. Examples of blowout jets are in Figure~1,
and in Extended Data Figures~2 and~3 and their corresponding videos (Extended Data Videos~2abc, 2def, 3abc
and 3def).  Examples  of standard jets are in Extended Data Figures~4 and~5 and their corresponding videos (Extended Data
Videos~4abc, 4def, 5abc and 5def). 


The flux-emergence model fails to explain the observation of a JBP occurring below the erupting minifilament,
which the Figure~2 picture naturally explains.  Also, an expectation of the emerging-flux model is that, as the
external reconnection progresses, reconnected open field will stand progressively closer to the JBP than that from
earlier reconnection\cite{moreno-insertis.et13}.  That is, the jet spire should drift {\it toward} the JBP with
time in that model.  Observations however show that more often than not the spire drifts {\it away} from the JBP
with time\cite{savcheva.et09}.  The schematic of Figure~2 naturally explains this proclivity for spire drift away
from the JBP, since the external reconnection of the erupting minifilament-carrying field produces reconnected
open field lines that in the corona stand progressively further away from the eruption's source location, which is
the location of the internal-reconnection flare arcade that is the JBP\@.

We have not addressed what leads to our minifilament eruptions.  Some recent studies of  on-disk coronal jets
found the miniature filaments to have likely resulted from cancelation of magnetic flux in the hours leading up to
the eruption\cite{shen.et12,adams.et14,young.et14}.  We suspect that, as with large-scale eruptions, various agents
could trigger the eruption, including flux cancellation and flux emergence.  For triggering by flux emergence, the
emergence would {\it trigger the minifilament's eruption,} rather than directly drive the jet as proposed
in the emerging-flux model for jets.

The minority-polarity flux in the base of the an X-ray jet presumably comes from flux emergence of compact field
loops into the dominant-polarity ambient field.  It therefore seems that many X-ray jets should be produced by
these closed-field emergences in the manner of the long-accepted emerging-flux model.  Our observation of no such
X-ray jets (at least for polar coronal holes) suggests that external  reconnection of the emerging closed field
with the ambient open field continually occurs fast enough to keep an appreciable current sheet from building up
at the magnetic null between the two fields, and a burst of enough external reconnection to make an X-ray jet can
be made only dynamically, driven by sudden eruption of the closed field as in a filament eruption.  That is, the
observed lack of emerging-flux-model X-ray jets suggests that no current sheet of the scale of the overall system
of two reconnecting fields can be formed gradually (i.e., quasi-stably) in the low-beta magnetised plasma of X-ray
jets, and by analogy nor in similar reconnection events in other low-beta astrophysical settings.




\clearpage


\begin{addendum} \item A.C.S. and R.L.M. were supported by funding from the Heliophysics Division of NASA's Science
Mission Directorate  through the Living With a Star Targeted Research and Technology Program (LWS TR\&T), and the Hinode
Project.  Both benefited from TR\&T discussions and from discussions with S. K. Antiochos.  ACS also benefited from
discussions held at the International Space Science Institute's (ISSI; Bern, Switzerland) International Team on Solar
Coronal Jets (Team Lead: N. Raouafi).

 \item[Author Contributions] A.C.S.: Reduction, analysis, and interpretation of XRT and AIA data; software development;
manuscript preparation. R.L.M.: Interpretation of results, manuscript review.  D.A.F.: Software development, assimilation and
calibration of AIA data.  M.A.: Discovery and analysis of seminal jet event motivating this broader investigation, manuscript
review.

 \item[Author Information] The authors declare that they have no
competing financial interests.  Correspondence and requests for materials should be addressed 
to A.C.S.~(email: alphonse.sterling@nasa.gov) or R.L.M. (ron.moore@nasa.gov).
\end{addendum}

\newpage

\begin{figure}
\hspace*{0cm}\includegraphics[angle=-90,scale=0.7]{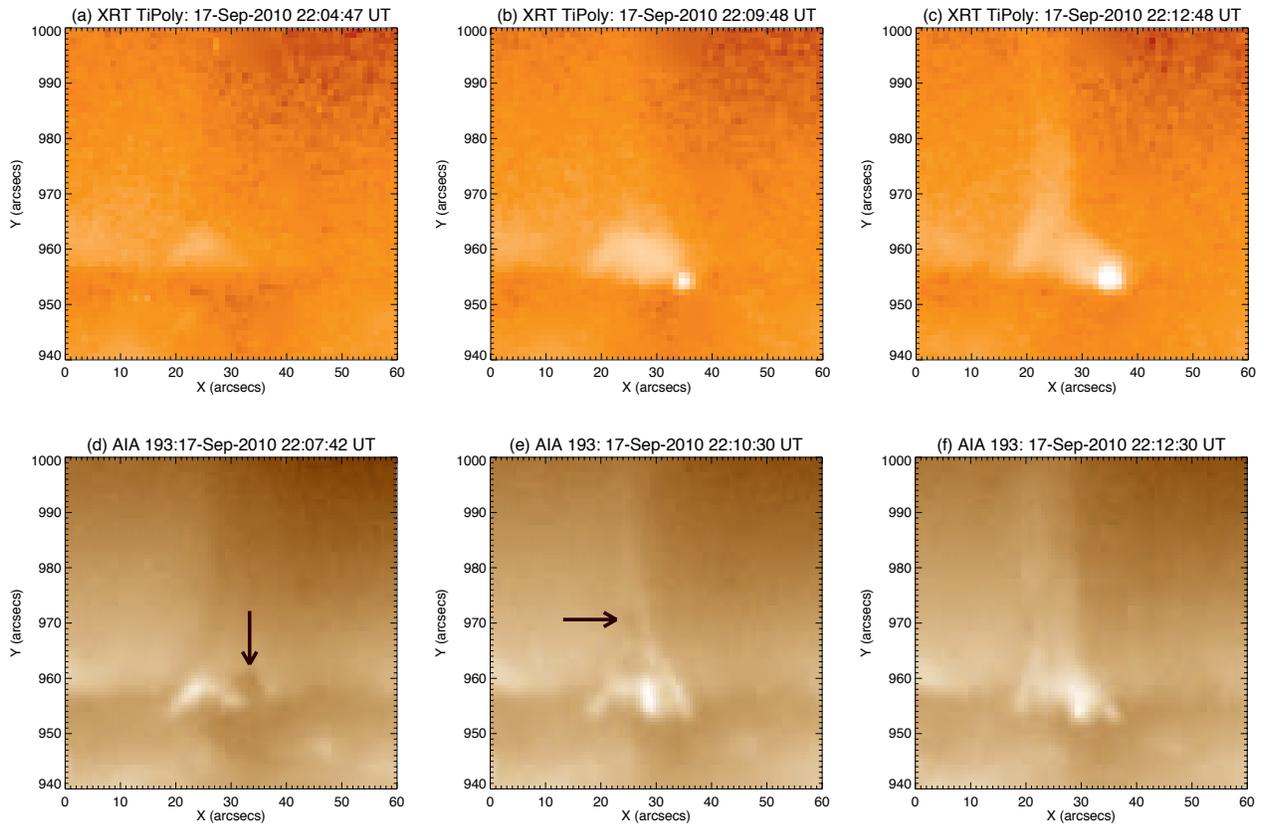}
\captionsetup{labelformat=empty}
\caption{{\bf Figure~1.  Erupting-Jet Example.}  Jet in soft X-rays (\hinode/XRT; a---c) and
EUV (\sdo/AIA 193~\AA; d---f). The jet bright point (JBP) is visible by the time of (b), and the jet is
fully developed and offset eastward of the JBP in (c).  Arrows show a minifilament moving outward
from the JBP location.  North  is upward and west is to the right.  Panels a and d, b and e, and c and f
are 217s, 30s, and 6s apart, respectively.  See Methods for details, and Extended Data for animations 
(Extended Data Video~1abc and 1def).  This is event~18 of Extended Data Table~1.}
\end{figure}
\clearpage

\begin{figure}
\hspace*{-0cm}\includegraphics[angle=0,scale=0.8]{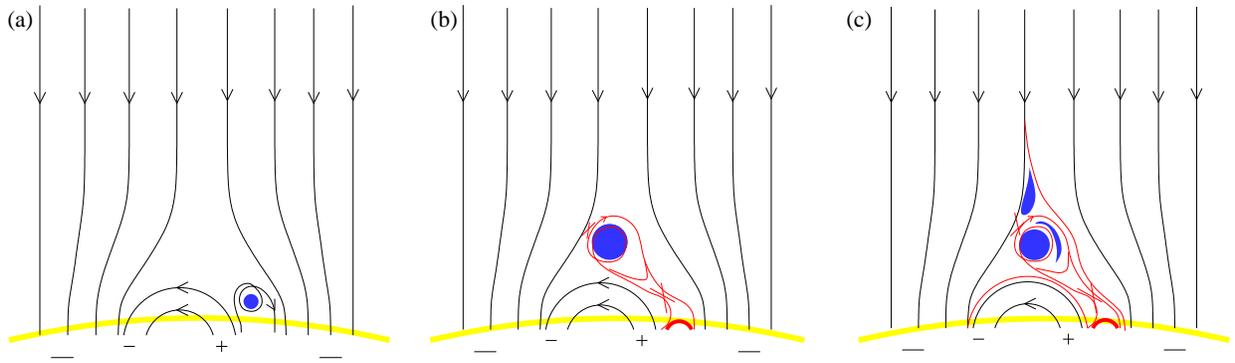}
\captionsetup{labelformat=empty}
\caption{{\bf Figure~2.  Revised Jet-Eruption Picture.}  Schematic representation of the
minifilament-eruption process that drives X-ray jets, as inferred from our observations.  Black lines
represent magnetic field, with arrows indicating polarities; red curves are newly-reconnected field
lines, blue features are minifilament material, and yellow curve is the solar limb.  From the initial
state (a), the jet forms as the minifilament erupts (b and c), with reconnection locations indicated by
red X-es (b and c).  The JBP (bold red arc) forms at the location of filament liftoff (b and c). See
Methods for more details.}
\end{figure}
\clearpage

\newpage
\begin{methods}

\subsection{Emerging-flux model.}  According to the emerging-flux
model\cite{yokoyama.et95,nishizuka.et08,moreno-insertis.et13,archontis.et13,fang .et14} (Extended Data~Fig.~1) for
solar coronal jets, an emerging bipole enters a dominant-polarity (negative in Extended Data~Fig.~1) ambient open
field, and the bipole's minority-polarity (positive) side can reconnect with coronal field  at the
location of the magnetic null region between the bipole and the ambient field.  After enough emergence of
the bipole, a burst of reconnection joins the outside of the bipole with nearby coronal field
(Extended Data~Fig.~1b), resulting in two reconnection products:  a small loop on the outside of the base of the
emerging bipole's minority-polarity side, and an open field connecting the bipole's majority-polarity
side with open coronal field, giving a new footpoint connection for that coronal field.  This type of
reconnection has been called ``interchange''\cite{crooker.et02}, or ``external'' \cite{sterling.et01},
since the reconnection is on the outside of the closed driving field (the emerging field in this case). 
An X-ray jet develops as reconnection-heated material flows out along the new open field strands. 
Additionally, the external-reconnection-formed small loop at the emerging field's edge is the model's explanation
for the observed base-edge compact X-ray jet bright point (JBP; also called a  ``hot loop''\cite{yokoyama.et95}). 
In the previous view of blowout jets, the idea was that the external reconnection causes and/or is driven by ejective 
eruption (blowout) of the emerging bipole, which is assumed to contain substantial non-potential 
(i.e.\ twisted) magnetic field, driving that bipole's eruption along the ambient open field to make a broad 
jet spire\cite{moore.et13}.

\subsection{Instrumentation and data.}

For our X-ray images, we use data from the \hinode/XRT with 30~s cadence and $1''$ pixels. XRT detects a broad
range of temperatures, but has highest sensitivity for temperature $T \gtsim 1.5$~MK, even for the relatively cool
``TiPoly''-filter images that we used for these observations.    For each jet in Extended Data Table~1 we studied
concurrent EUV images from the SDO/AIA, with $0''.6$ pixels and 12~s cadence. Our final movies and figures were
formed by summing the frames in pairs, and therefore the resulting movies we used were generally of 1-min cadence
and 24-sec cadence respectively for XRT and AIA\@.  This summing blurs the images somewhat, but renders subtle
features, such as X-ray jets and some of the fainter EUV minifilaments, much easier to discern.  For many of the
X-ray jets of our study, we examined all the AIA EUV channels, which are tuned to wavelengths of 304~\AA, 171~\AA,
193~\AA, 211~\AA, 131~\AA, 335~\AA, and 94~\AA; these respectively have strong responses to log temperatures (K)
of: $\approx$~4.7, 5.8, 6.2, 6.3, 7.0,  6.4, and 6.8 (although some channels are multivalued\cite{lemen.et12}). 
Usually there was little new information in the  hotter 131~\AA, 335~\AA, and 94~\AA\ channels, and so we did not
inspect these hotter channels for some of the jets.

In total we examined 20 X-ray jets, initially selected during an earlier study\cite{moore.et13}, where the
JBP was obvious in the X-ray images (Extended Data Table~1).  From the previous
study\cite{moore.et13}, each event of Extended Data Table~1 is typed as ``standard'', ``blowout'', or
``ambiguous'', based on its morphology in the XRT images (and in some  cases, in AIA 304~\AA\ images also).
Blowout jets are those where the entire base brightened and where the spire broadened with time to span
approximately the width of the base, while standard jets are those where only the JBP
brightened substantially in the base and the spire remained narrow compared to the span of the base.  The JBP is
also referred to by other terms, including ``hot loop''\cite{yokoyama.et95}, ``bright
loop''\cite{yokoyama.et95},   ``bright point''\cite{yokoyama.et95,moore.et10,moore.et13}, and ``bright
footpoint''\cite{savcheva.et07}.

In each blowout jet in Extended Data Table~1 the minifilament eruption was evidently ejective; the erupting
closed field apparently blows out into the ambient open field.  In this case, much or all of the filament
material escapes from the closed field onto the open field.

In the events of Extended Data Table~1 categorised as standard jets, a minifilament eruption was detectable, but usually
that eruption appeared either not be ejective, or it was perhaps ejective but weak and/or  faint.  In event 4, a
minifilament (best seen in 304~\AA) has a whipping motion from the location that becomes the JBP\@. Event~7 seems to be
generated by a minifilament that becomes partially destabilized and spins (rolls) beneath confining magnetic fields. 
These standard-jet events therefore may be analogous to larger-scale confined filament eruptions, ones that make flares
that are of shorter duration than the ejective flares\cite{moore.et01}. (As an example, the rolling minifilament of
Event~7 could be a scaled-down version of the confined filament  eruption in Fig.~1 and in the corresponding on-line
movies of the paper Sterling~\etal~2011\cite{sterling.et11}.) Event 6, another standard jet, however shows an ejective
minifilament, similar to the jets identified as blowout jets, but it does not make a broad spire.  In that case it appears
as if the minifilament erupted far enough for much of it to escape onto open field via external reconnection, but not 
enough to blow out violently and make a broad jet.  In comparison to the blowout jets, more of the filament material remains
trapped within the closed field.

Our other standard jets (5, 19), and the ambiguous jet (11), similarly may be partially-confined and  partially-ejective
minifilament eruptions. In these cases, some of the minifilament material escapes onto the  open field, and some of it
remains in the closed field.  In this sense, we envision a continuum of jets manifestations, between pure blowout jet
(where the filament field would push far into the opposite-polarity open field, making a broad jet, and all of the
filament material would eventually escape onto that open field), and a pure standard jet (where only the envelope of the
closed filament field reconnects with the opposite-polarity open field, and none of closed field containing the cool
filament material undergoes external reconnection). Our view of standard jets being due to confined
minifilament eruptions, partially-confined minifilament eruptions, and/or weak ejective minifilament eruptions, is still
speculative.  Further study will be required to understand fully various morphological differences among jets.

\subsection{Minifilament measurement details.}

We measured the line-of-sight lengths and velocities of the minifilaments during the period after they started
to erupt but prior to when they formed a jet or reached the apex of the base (below the jet spire). We usually 
used the 171~\AA, 193~\AA, or
211~\AA\ AIA channels for these measurements; only for  events 4, 7 and 10 did we find 304~\AA\ preferable for
determining minifilament properties in our  data set.  We obtained mean velocities for the erupting
minifilaments of $31 \pm 15$~\kms; if the velocities are weighted inversely with their uncertainty (Extended
Data Table~1), the weighted mean velocity and weighted standard deviation are 24~\kms\ and 13~\kms,
respectively.

\subsection{Details of jet-formation process in our picture.}

As shown in Fig.~2, we envision that initially a minifilament-carrying, non-potential, relatively-compact core  field
of a magnetic bipole (or magnetic arcade) sits next to (and shares the minority-polarity flux with) a relatively-large 
bipole (Fig.~2a).  An unspecified process
destabilizes the smaller bipole so that it erupts, with the minifilament being  channeled between the large bipole and
the overlying open field.  Upon reaching the open  coronal field on the far side of the large bipole, the field
carrying the  minifilament reconnects with that field (Fig.~2b), and a jet, often  including substantial minifilament
material, is ejected  along newly-reconnected open field (Fig.~2c).  This reconnection also adds field to the large
bipole. Also in (Fig.~2b), internal  reconnection (lower red X) of the minifilament-carrying field
occurs; this is reconnection internal to the erupting lobe of the double bipole, and this reconnection forms a
flare arcade (the JBP) in the wake of the ejected minifilament.

This is analogous to the formation of commonly-observed flare arcades in typical large-scale solar eruptions; that is, the
erupting lobe of the system erupts as in a ``typical'' large-scale eruption, as pictured in, e.g., Figure~1 of
Shibata~\etal~(1995)\cite{shibata.et95} or Figure~1 of  Moore~\etal~(2001)\cite{moore.et01}.  In our jet-formation picture,
this process is occurring on a smaller scale so that the filament of those typical models corresponds to our  minifilaments. 
Rather than a filament traveling directly outward as in those schematics of large-scale eruptions, in the jet case of Figure~2
the minifilament travels along the curved path between the adjacent bipole and distorted ambient coronal field.  (The coronal
field is distorted by the magnetic field of the two bipoles.) As long as the erupting minifilament is on the near-side (i.e.,
the side of its origin) of the apex of the neighboring bipole, no reconnection occurs between the erupting-bipole field
enveloping the filament and ambient coronal fields.  (In 3D the situation will not be as pure as in the 2D schematic, but we
still expect the basic situation of the 2D schematic to hold.)  We will consider what happens when the enveloping field reaches
the far side of the apex shortly.  First however, again looking at the schematics of the typical large-scale eruptions, it can
be seen that the field lines beneath the erupting filament reconnect (this is what we are calling internal reconnection, since
it occurs internal to the erupting bipole) to form hot flare loops near the solar surface. In our analogous schematic for the
jets in Figure~2, these flare loops correspond to the JBP\@.  While the small lobe of the double bipole in Fig.~2 is erupting
in this fashion, the neighboring bipole largely remains inert, except for the addition of the new field via the external
reconnection, as mentioned in the previous paragraph. 

Now consider the situation when the erupting-minifilament bipole reaches the far side of the apex of the
neighboring bipole (Fig.~2b).  Because the field orientations are then opposite, the erupting field enveloping the minifilament
and the far-side ambient coronal field can undergo reconnection; since this reconnection is between the field of the
erupting bipole and coronal field that is external to that erupting bipole, we call this external reconnection.  This
external reconnection adds heat to the reconnected field lines, making a hot jet spire along the open field lines and
making hot loops over the adjacent bipole (red curves in Fig.~2c).  This external reconnection progressively
erodes away the field enveloping the cool minifilament material.  If this erosion of that enveloping field stops before
the field lines holding cool material is reached---which could happen, for example, if the erupting minifilament-carrying
bipole does not have enough energy to travel deep into the far-side ambient-field region---then the cool material never
makes it onto the open field (and the spire receives no cool material).  Rather, the filament plasma remains trapped in 
closed field in the base of the jet.  This  may be how the standard jets are formed; only a narrow
hot spire forms if the erupting minifilament-carrying bipole does not go far into the ambient-field region.

In a  blowout jet, the eruption continues deeper into the opposite-directed ambient field region to make a broader spire
than is depicted in Fig.~2c. The envelope around the cool-minifilament material is completely eroded away, and so the
cool material escapes onto the open ambient coronal field, forming a cool jet.  In this sense, the eruption of the
minifilament is analogous to ejective eruptions of typical large-scale cases. (Some standard jets appear to be weak
versions of such ejective jets.)   The drawings in Fig.~2 are tailored to depict the jet in Fig.~1, jet~18 in Extended
Data Table~1, which is a blowout jet.

The external reconnection of the erupting-minifilament field with the open field also adds a new hot layer to the
larger  bipole (larger red loop in Fig.~2c); this reconnection product from earlier eruption episodes might have
created the ``initial'' large  bipole (large black loops of Fig.~2a). Other possibilities for the initial large
bipole are that it, along with with the filament-carrying bipole, are two asymmetric lobes of an anemone field
region\cite{shibata.et07}.  That anemone region could be due to recently-emerged magnetic flux, or it could have
formed over time via surface-flux migration and cancelation\cite{innes.et13}.

A schematic for X-ray jets similar to that of our Figure~2 was presented by Shibata~(1999)\cite{shibata99}, in Figure~8b
of that paper. That figure was derived from data from earlier satellite missions, prior to the high resolution,
high-cadence, multiple-EUV-wavelength data of SDO/AIA\@.  There is however a difference between that picture for jets and
our picture.  The proposal there\cite{shibata99} is that a plasmoid (which might correspond to our minifilament) erupts
from the external-reconnection site of the  emerging-flux model (Extended Data Fig.~1), the pre-eruption plasmoid being
in the current sheet between the emerging flux and the ambient coronal field. (Also,  Figure~6 of that same
paper\cite{shibata99} explicitly depicts an emerging-flux origin for X-ray jets.) In contrast, our proposal is that X-ray
jets, at least in coronal holes, are a miniature version of the standard model for large-scale flares and CMEs, independent 
of whether there is emerging flux.  In our view, prior to eruption the
minifilament resides in sheared field (or in a flux rope) in the core of a magnetic arcade, instead of in  a current
sheet.  More  generally, in our view the triggering and eruption of the minifilament may include any of  a multitude of
processes and subprocess proposed for large-scale eruptions, including those listed in the main text, and
others\cite{lin.et01,shibata.et01,chen11,kusano.et12}.  Determining whether the pre-eruption minifilaments that erupt in
jets are  located at an external-reconnection current sheet as suggested by Shibata~(1999)\cite{shibata99}, or instead
reside in a magnetic arcade as we picture, requires further observational study.

In AIA movies the developing jets show clear rotation in some cases, such as the jet of Figure~1  (Extended Data
Video~1def).  Other jets however show only partial rotation (e.g., Extended Data Figure~2, and Extended Data
Videos~2abc and~2def), or no obvious rotation (e.g., Extended Data Figure~3, and Extended Data Videos~3abc 
and~3def). Since we have not identified a clear pattern regarding the rotations and the resulting jets, we  do not
address this topic further here.

\subsection{Cause of minifilament-eruption onset.}

Since in this study we do not examine jets that originate at low solar latitudes, we cannot adequately see
the causes (triggering) of these magnetic eruptions.  As with large-scale filament eruptions, several
triggering agents could be responsible, including flux cancelation or even flux emergence.   Our main point
here is that, independent of the cause of the minifilament-eruption onset, the jets all result from those
minifilament eruptions, with the JBP being  the ``flare'' occurring in conjunction with those minifilament
eruptions.

As stated in the main text however, several other studies\cite{shen.et12,adams.et14,young.et14} found
on-disk coronal jets clearly to occur in conjunction with magnetic flux cancelation. 
One study\cite{adams.et14} aggressively searched for emerging flux beneath a jet, but found no significant
signature of emergence.  A different study\cite{chandrashekhar.et14} also searched for but did not find
emerging flux below an on-disk coronal jet. Another study\cite{innes.et09} found mini-CMEs resulting from
perhaps ``small filament ejections'', that may be similar or identical to the jets we discuss here; they
report the ejections to occur at sites of  ``twisting small concentrations of opposite polarity magnetic
field'', and again they do not report detections of emerging flux.  Similar jets were reported
elsewhere\cite{schrijver10}, but without direct magnetic field observations.

We have found two studies of on-disk jets where emerging flux was reported.  In the first of these\cite{liu.et11},
although emergence occurred, a microflare and an EUV jet happened only after cancelation of flux in the region of
the flux emergence. Similarly, in the second study\cite{huang.et12} flux emergence occurred, but, for two
different jets they observed, both jets occurred at about the time the emerged flux underwent cancelation with
neighboring field.  In that second case, the jet observations were from XRT, and were of jets occurring in on-disk
coronal holes; thus those observations are on-disk complementary examples of the near-limb XRT
polar-coronal-hole jet observations we present in this paper.

On balance then, the on-disk coronal jet studies suggest that flux cancelation is often crucial to jet
onset.  In light of the present study, we expect that in those earlier observations, the cancelation likely
resulted in minifilament eruptions that produced jets, with flares occurring in the wake of those eruptions
and appearing as JBPs.

\clearpage
\centerline{Additional References (see main text for 1-30)}

\let\oldthebibliography=\thebibliography
\let\oldendthebibliography=\endthebibliography
\renewenvironment{thebibliography}[1]{%
    \oldthebibliography{#1}%
    \setcounter{enumiv}{30}%
}{\oldendthebibliography}

\end{methods}


\newpage

\noindent
{\bf Extended Data Follow}

\captionsetup[table]{name=Extended Data Table,labelfont=bf,textfont=bf}

\begin{table}
\caption{X-Ray Jets of This Study}
\renewcommand{\arraystretch}{0.6}
\tabcolsep=0.11cm
{\footnotesize
\begin{tabular}{ccccccc} \hline
Event & Date$^{a}$ & Start; End$^{b}$ & $x,y$ (arcsec)$^{c}$ & Type$^{d}$ & Fil.~Size$^{e}$ (arcsec) &
Fil.\ Speed$^{e}$ (\kms)\\ 
\hline
 1 & 2010 Jul 24 & 15:56; $>$16:15 & -60, 950 & blowout & 17 &$14 \pm 2$ \\	
 2 & 2010 Jul 25 & 12:29; 12:46 & 140, -950 & blowout & 10 & $30\pm10$ \\	
 3 & 2010 Aug 26 & 14:13; $>$14:16 & 100, 950 & blowout & 10 & $28 \pm 5$ \\	
 4 & 2010 Jul 27 & 11:35; 12:17 & 30, 920 & standard & $20$$^{f}$ &$50 \pm5$$^{f}$ \\	
 5 & 2010 Jul 27 & 11:40; 12:20 & -50, 920 & standard  & diffuse(?)$^{f}$ 
                                           &$28 \pm 5$(?)$^{f}$\\	
 6 & 2010 Aug 28 & 11:40; 12:03 & -130, 940 & standard & $5 $ & $28 \pm 5$ \\	
 7 & 2010 Aug 28 & $<$13:41; $>$13:48 & -70, 840 & standard & 17 & rolling\\    
 8 & 2010 Sep 05 & 21:14; 21:35 & 30, 840 & blowout & 10 & $28 \pm 5$ \\	
 9 & 2010 Sep 08 & 01:29; 01:44 & 40, 935 & blowout & $6$ & $19 \pm 5$ \\	
 10 & 2010 Sep 09 & 20:14; 20:33 & 20, 770 & blowout & 17 & $73 \pm 8$ \\	
 11 & 2010 Sep 09 & 20:21; 20:40 & 60, 850 & ambiguous &  12 
                                         &  uncertain$^{g}$ \\	
 12 & 2010 Sep 09 & 22:05; 22:31 & 0, 910 & blowout &  7 & $13\pm 3$ \\		
 13 & 2010 Sep 09 & 23:52; 00:06 & -120, 950 & blowout & 9 & $33\pm 5$ \\	
 14 & 2010 Sep 10 & 00:01; 00:09 & -10, 880 & blowout & 7 & $50 \pm 10$\\	
 15 & 2010 Sep 11 & 00:39; 00:50 & 80, 950 & blowout & 8$^{f}$ & $19 \pm 5$$^{f}$\\	
 16 & 2010 Sep 11 & $<$01:08; 01:27 & -120, 950 & blowout & 13 & $40 \pm 8$\\	
 17 & 2010 Sep 17 & 20:39; 21:08 & -20, 840 & blowout & diffuse$^{h}$ 
                                           & $33 \pm 8$$^{h}$ \\	
 18 & 2010 Sep 17 & 22:08; 22:18 & 30, 960 & blowout & 7 & $40 \pm 5$\\		
 19 & 2010 Sep 19 & 19:47; 20:23 & 20, 880 & standard & 10 & $20\pm 5$\\	
 20 & 2010 Sep 27 & 00:39; 00:43 & 0, 960 & blowout & 10 & $20\pm 5$\\		
\hline 
\end{tabular}

\vspace{0.6cm}
\noindent
(a) Date of event start time. \\
(b) UT time period of clearly-detectable jet and/or compact X-ray jet bright point (JBP) in XRT images. Symbols ``$<$''
   and ``$>$'' respectively indicate the jet started before or continued after indicated times during gaps in XRT data.\\
(c) Approximate $x,y$ location of jet in AIA images in heliocentric coordinates. \\
(d) Morphological classification of X-ray jet based on Moore~\etal\cite{moore.et13}\ study. \\
(e) Line-of-sight projected size/speed of minifilament near time of eruption onset; size uncertainty $\ltsim 3''.$ \\
(f) Minifilament diffuse, faint, or identification less certain than other cases. \\
(g) Accurate speed measurement not possible due to image shifts during eruption time. \\
(h) Minifilament too diffuse for size measurement, but moving structures can be tracked for velocity estimate. \\
} 
\end{table}

\clearpage

\newpage
\begin{figure}
\hspace*{1.5cm}\includegraphics[angle=0,scale=0.7]{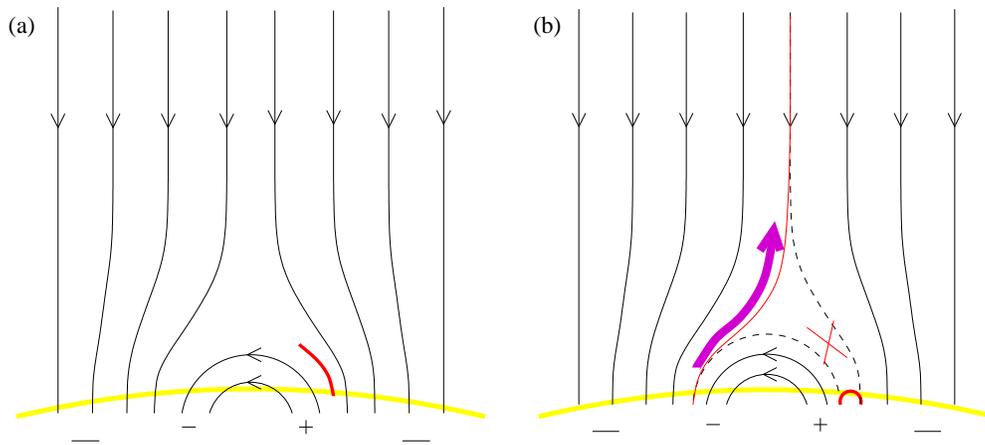}
\captionsetup{labelformat=empty}
\caption{{\bf Extended Data Figure~1.  Emerging-Flux Model for Solar Jets.} Schematic representation of
the commonly-accepted solar X-ray-jet-formation mechanism\cite{shibata.et92}. Lines and red cross are as
in Fig.~2, and the red curve in (a) represents a current sheet.  Flux emergence purportedly forces
reconnection at the current sheet (red cross), resulting in new closed loop field (red loop), and new
connections to the open coronal field (red curve), along which the X-ray jet (magenta) flows.  According
to this model, the new reconnection loops appear as the JBP\@.  Previous scenarios for 
``blowout jets''\cite{moore.et10,moore.et13,pariat.et09} have been variations of this model.}
\end{figure}
\clearpage

\newpage
\begin{figure}
\hspace*{0cm}\includegraphics[angle=-90,scale=0.7]{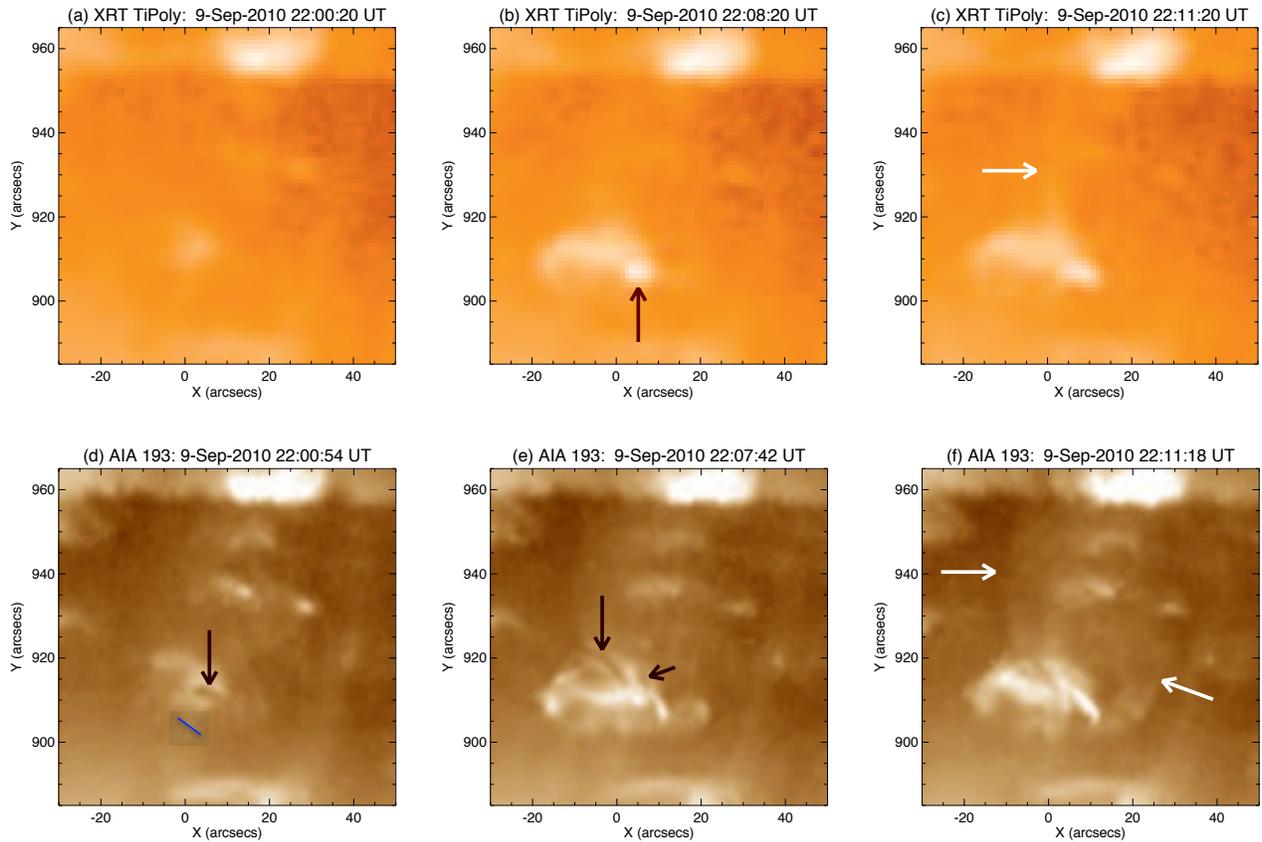}
\captionsetup{labelformat=empty}
\caption{{\bf Extended Data Figure~2. Extended Data Table~1 Event~12 Jet.} XRT (a---c) and AIA~193\AA\
(d---f)  images of the jet. Arrows show: (b) developing JBP; (c) X-ray-jet spire; (d) minifilament; in (e)
both arrows point to segments of the minifilament, which split during eruption; (f) both arrows point to edges
of a broad jet. In (d) the blue bar shows our estimate of the size of the minifilament, the value of which
appears in Extended Data Table~1.  See Extended Data Video~2abc and 2def for animations.}
\end{figure}
\clearpage

\newpage
\begin{figure}
\hspace*{0cm}\includegraphics[angle=-90,scale=0.65]{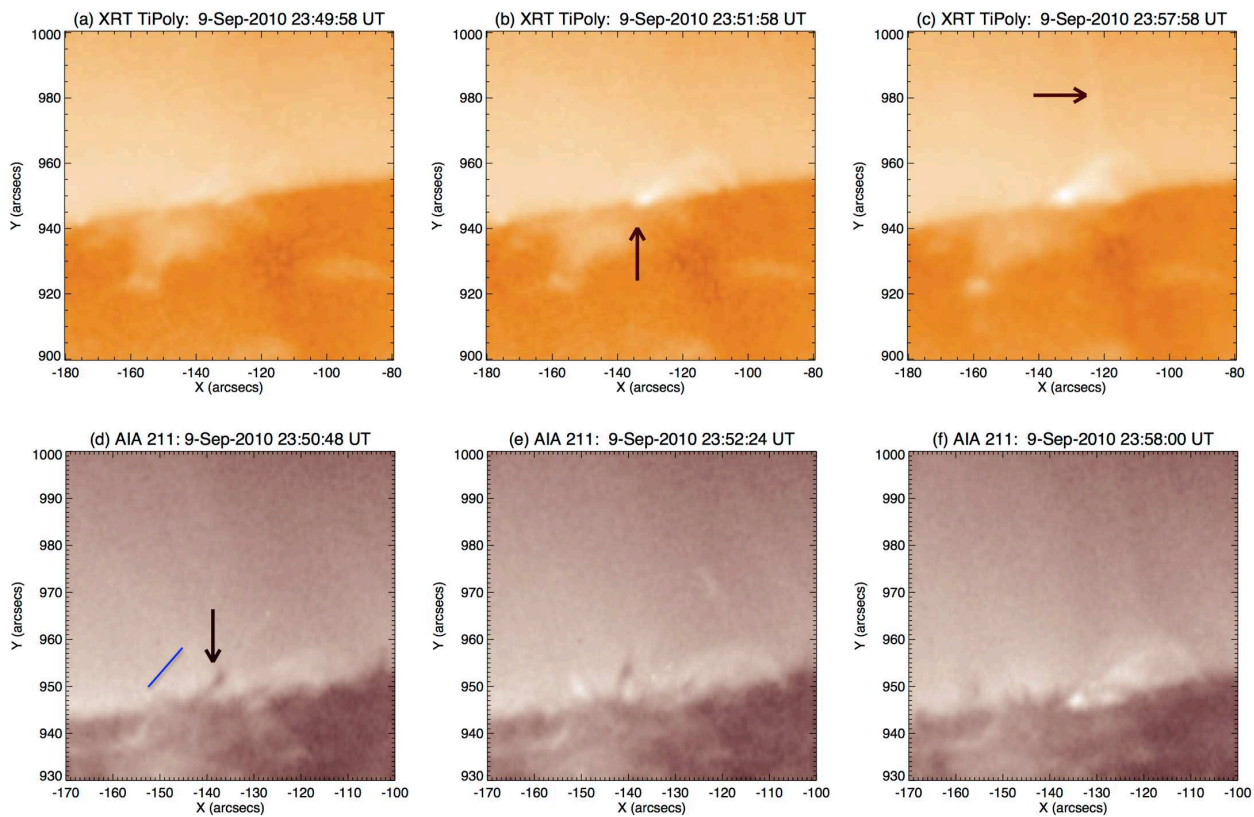}
\captionsetup{labelformat=empty}
\caption{{\bf Extended Data Figure~3. Extended Data Table~1 Event~13 Jet.} 
XRT (a---c) and AIA~211~\AA\ 
(d---f) images of the jet. Dark feature in upper-right of XRT images is an artifact. Arrows show: (b) developing JBP; (c)
X-ray-jet spire; (d) minifilament starting to erupt.  Blue  bar in (d) is as in Extended Data Fig.~2.   AIA images
are zoomed-in more than XRT images.  See Extended Data Video~3abc and 3def for  animations.}
\end{figure}
\clearpage

\newpage
\begin{figure}
\hspace*{0cm}\includegraphics[angle=-90,scale=0.7]{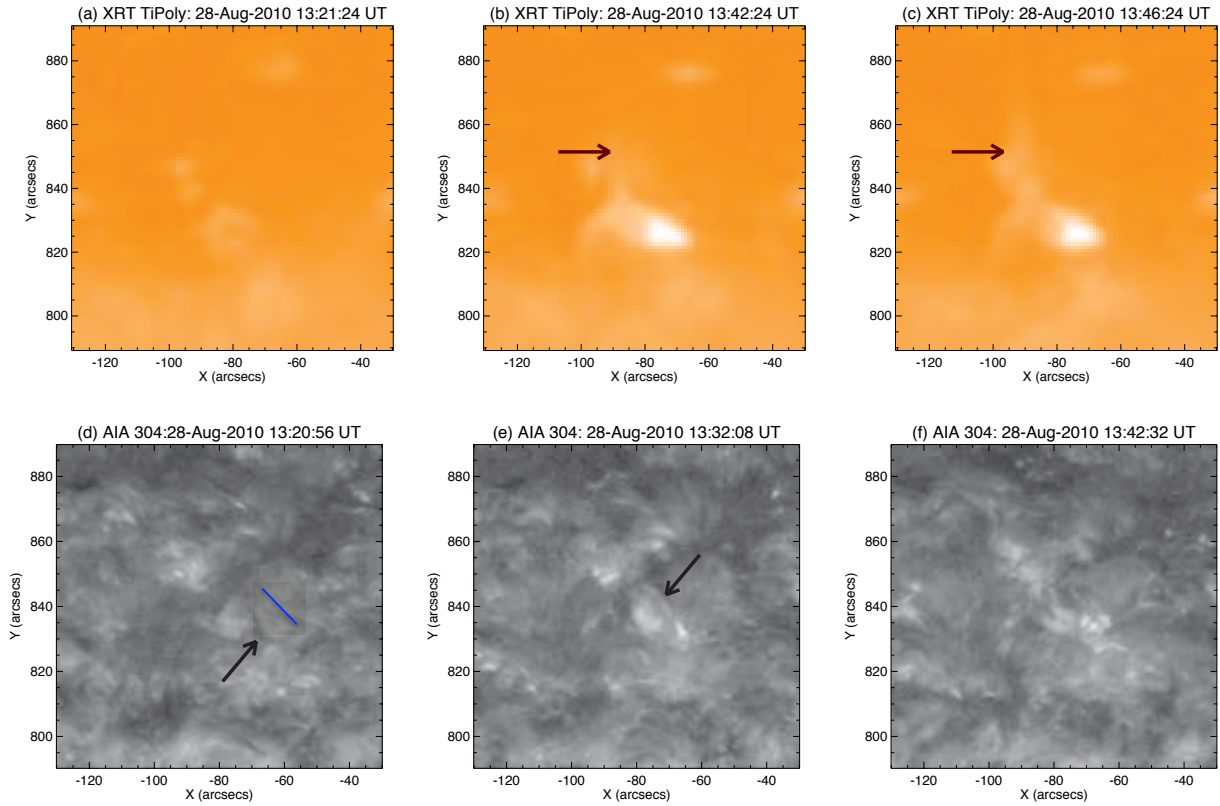}
\captionsetup{labelformat=empty}
\caption{{\bf Extended Data Figure~4. Extended Data Table~1 Event~7 Jet.} XRT (a---c) and AIA~304\AA\ (d---f)
images of a ``standard'' jet.  Arrows show: (b) X-ray jet spire; (c) X-ray jet spire, showing drift since (b);
(d) minifilament starting to erupt; (e) ``rolling'' filament (see Methods).  Blue  bar in
(d) is as in Extended Data Fig.~2.  Grey-scale (d---f) shows filament better than colour for this event. See
Extended Data Video~4abc and 4def for animations.}
\end{figure}
\clearpage

\newpage
\begin{figure}
\hspace*{0cm}\includegraphics[angle=-90,scale=0.7]{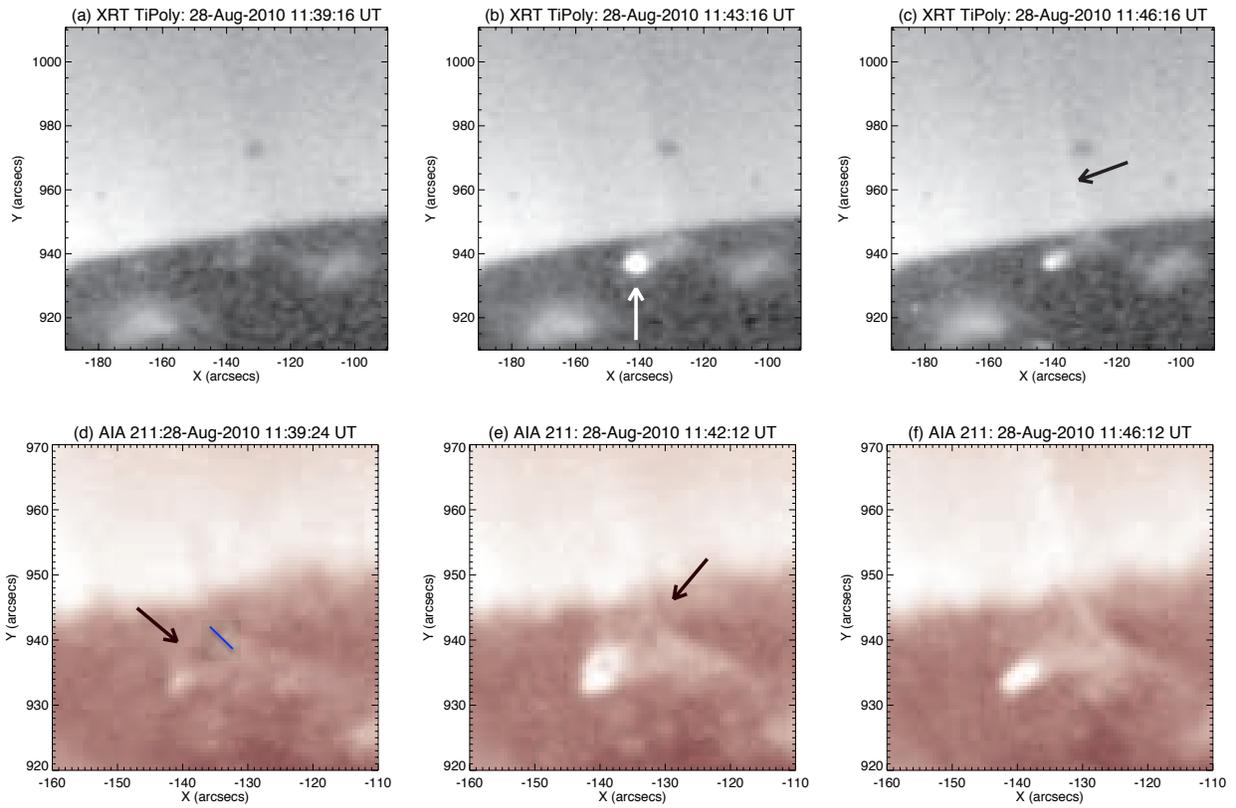}
\captionsetup{labelformat=empty}
\caption{{\bf Extended Data Figure~5. Extended Data Table~1 Event~6 Jet.} XRT (a---c) and AIA~211\AA\ (d---f)
images of a ``standard'' jet.  Dark spot north-west of center in XRT images is an artifact.  Arrows show: (b) JBP;
(c) X-ray jet spire; (d) minifilament moving upward; (e) minifilament near apex of jet base, with jet spire
starting to develop.  AIA images are zoomed-in more than XRT images.   Blue  bar in (d) is as in Extended Data
Fig.~2.  See Extended Data Video~5abc and 5def for animations.}
\end{figure}
\clearpage

\end{document}